\documentclass[conference]{IEEEtran}

\usepackage{cite}
\usepackage{float}
\usepackage{graphicx}
\usepackage{amsmath}
\usepackage{times}
\usepackage{graphicx}
\usepackage{latexsym}
\usepackage{graphicx}
\usepackage{bm}
\usepackage{amssymb}
\usepackage[center]{caption2}
\usepackage{stfloats}
\usepackage{cases}
\usepackage{array}
\usepackage{setspace}
\usepackage{fancyhdr}
\usepackage{color}
\usepackage{subfigure}
\usepackage{multirow}
\usepackage{amsmath}
\usepackage{cases}
\usepackage[noend]{algpseudocode}
\usepackage{algorithm}
\usepackage{algorithmicx}
\usepackage{diagbox}
\usepackage{algpseudocode}
\usepackage{epstopdf}
\usepackage{varwidth}
\usepackage{pifont}
\usepackage{mathrsfs}
\usepackage{booktabs} 

\usepackage{verbatim}

\usepackage[colorlinks,urlcolor=blue,linkcolor=black]{hyperref}

\usepackage{url}

\begin{document}

\title{AMP-SBL Unfolding for Wideband MmWave Massive MIMO Channel Estimation}



\author{\IEEEauthorblockN{Jiabao Gao\IEEEauthorrefmark{1},
		Caijun Zhong\IEEEauthorrefmark{1}, Geoffrey Ye Li\IEEEauthorrefmark{2}}
	    \IEEEauthorrefmark{1}College of Information Science and Electronic Engineering, Zhejiang University\\
	    Hangzhou, China, E-mail: \{gao\_jiabao, caijunzhong\}@zju.edu.cn\\
		\IEEEauthorrefmark{2}Department of Electrical and Electronic Engineering, Imperial College London\\
		London, UK, E-mail: geoffrey.li@imperial.ac.uk
	}
	
\maketitle

\begin{abstract}
In wideband millimeter wave (mmWave) massive multiple-input multiple-output (MIMO) systems, channel estimation is challenging due to the hybrid analog-digital architecture, which compresses the received pilot signal and makes channel estimation a compressive sensing (CS) problem. However, existing high-performance CS algorithms usually suffer from high complexity. On the other hand, the beam squint effect caused by huge bandwidth and massive antennas will deteriorate estimation performance. In this paper, frequency-dependent angular dictionaries are first adopted to compensate for beam squint. Then, the expectation-maximization (EM)-based sparse Bayesian learning (SBL) algorithm is enhanced in two aspects, where the E-step in each iteration is implemented by approximate message passing (AMP) to reduce complexity while the M-step is realized by a deep neural network (DNN) to improve performance. In simulation, the proposed AMP-SBL unfolding-based channel estimator achieves satisfactory performance with low complexity.
\end{abstract}



\section{Introduction}

Millimeter wave (mmWave) massive multiple-input multiple-output (MIMO), which enjoys the benefits of huge bandwidth\cite{mmWave} and spatial degree of freedom\cite{massive_mimo}, is recognized as an indispensable technology to meet the high data rate requirements of future wireless communication systems. To reduce hardware cost and power consumption, the hybrid analog-digital architecture needs to be adopted, where massive antennas are connected to only a few radio frequency (RF) chains through a phase shifter network\cite{had}. 

To achieve theoretical gains of massive MIMO, accurate channel state information is essential, whose estimation, however, is challenging in the hybrid architecture. Since the overhead of linear estimators increases dramatically, compressive sensing (CS) algorithms are usually preferred, where the high-dimensional channel is directly recovered from the compressed received pilot signal. In existing CS-based channel estimators, the channel sparsity in the angular domain caused by the limited scattering characteristic of mmWave signals is leveraged by default, while the dimension of subcarriers can be handled in two different ways, leading to two categories of estimators. The first category is the angular-frequency (AF) estimator, where the common angular sparsity structure among all subchannels is exploited to improve the estimation accuracy in the angular domain. Specifically, in \cite{somp_ce}, the simultaneous orthogonal matching pursuit (SOMP) algorithm is proposed, where the projections on all subcarriers are averaged to reduce the equivalent noise. Similar ideas to embed the common angular sparsity structure also apply to the approximate message passing (AMP) algorithm\cite{mmv_amp} and the sparse Bayesian learning (SBL) algorithm\cite{sbl_ce} for performance improvement. Different from the AF estimator, the angular-delay (AD) estimator further transforms the frequency domain subchannels into the sparse delay domain and exploits the double channel sparsity in the AD domain. In \cite{amp_3d}, AMP with nearest neighbor pattern learning is proposed to exploit the block sparsity structure in the AD domain, while other CS algorithms like iterative shrinkage thresholding (ISTA)\cite{beam_squint_LISTA} can be used as well. In spite of having better performance than AF estimators, the practical applications of AD estimators are hindered by their high complexity, since the sparse transformation from the frequency domain to the delay domain results in sparse vectors and measurement matrices of larger dimensions. 

In the wideband mmWave scenario, a unique challenge of massive MIMO channel estimation is the non-negligible beam squint effect caused by the combination of huge bandwidth and massive antennas, which destroys the common angular sparsity structure among subchannels in AF estimators. In AD estimators, beam squint will lead to energy diffusion and reduce the AD channel sparsity\cite{wideband_effect}. To deal with beam squint, a special detection window is proposed in \cite{dll_window} to capture the beam squint pattern instead of simply averaging variables from different subcarriers. Frequency-dependent angular dictionaries are adopted in \cite{gff_block} so that the common angular sparsity still holds with beam squint. In \cite{beam_squint_ce}, the majorization-minimization iterative approach is used, which naturally incorporates general channel effects including beam squint.

Recently, deep learning (DL) has achieved great success in physical layer wireless communications thanks to its strong representation ability. For wideband mmWave channel estimation with beam squint, \cite{beam_squint_hht} proposes to realize the denoiser in the  generalized expectation consistent (GEC) algorithm by a deep neural network (DNN), so that the beam squint patterns on the channel image can be handled properly. To fully exploit channel sparsity, denoising is further performed in the learned sparse transform domain in \cite{beam_squint_LISTA}, which is similar to AD estimators. In our previous work\cite{sbl_unfolding_ce_frequency}, the expectation-minimization (EM)-based SBL algorithm is unfolded. In each iteration, the update rule of Gaussian variance parameters in the M-step is learned by a tailored DNN, which can learn the non-ideal common angular sparsity structure. In \cite{sbl_learned_epsilon}, only the scalar shape hyperparameter is adaptively tuned by the DNN in each SBL iteration, which can be regarded as a special case of \cite{sbl_unfolding_ce_frequency} with fewer degrees of freedom. 

Our previous work\cite{sbl_unfolding_ce_frequency} still belongs to the AF estimator. To achieve more accurate channel estimation, in this paper, we extend the SBL unfolding approach to the AD estimator, where the design of DNN-based M-step is inherited while two modifications are made to enhance both performance and complexity. On one hand, frequency-dependent angular dictionaries are used to avoid energy diffusion and promote AD channel sparsity. On the other hand, the complexity of the algorithm is dramatically reduced by implementing the E-step with the AMP algorithm. According to simulation results, the proposed approach can achieve accurate channel estimation with low complexity.

\section{System model and problem formulation}
In this section, we first present the hybrid massive MIMO system and the pilot signal transmission model. After introducing the wideband mmWave channel model, the channel estimation is formulated as a CS problem.
\begin{figure}[!htb]
	\centering
	\includegraphics[width=0.5\textwidth]{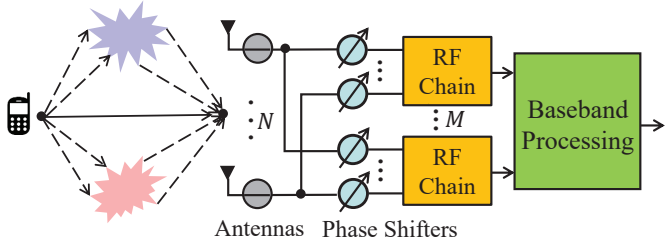}
	\caption{Hybrid massive MIMO system.}
	\label{system}
\end{figure}

\subsection{System Model}
As illustrated in Fig. \ref{system}, we consider a hybrid massive MIMO system where the base station (BS) equipped with an $N$-antenna uniform linear array (ULA) and $N_{RF}$ RF chains serves a single-antenna user. The extension to the multi-antenna multi-user scenario is also straightforward by allocating orthogonal resources to different users and user antennas. To deal with frequency selective channels, the total system bandwidth $f_s$ is evenly divided by $K$ subcarriers. In the $q$-th channel use, the received signal at the BS at the $k$-th subcarrier can be expressed as
\begin{equation}
	\bm{y}_{q}^k=\bm{W}_q(\bm{h}^ks_{q}^k+\bm{n}_{q}^k)\in \mathbb{C}^{N_{RF}\times 1},
\end{equation}
where $s_{q}^k$, $\bm{h}^k\in \mathbb{C}^{N\times 1}$, $\bm{n}_{q}^k\sim \mathcal{CN}(\bm{0},\sigma^2\bm{I}_{N})$, and $\bm{W}_{q}\in \mathbb{C}^{N_{RF}\times N}$ denote the transmitted pilot signal at the user, the channel vector, the noise vector before combining with variance $\sigma^2$, and the receive combining matrix, respectively. Since the pilot signal is known at the BS, we set $s_{q}^k=1,\forall q,k$ for convenience, and define the transmit signal-to-noise-ratio (SNR) as $1/\sigma^2$. Concatenating the received signals of totally $Q$ channel uses, we have
\begin{equation}
	\bm{y}^k=[(\bm{y}_{1}^k)^T,\cdots,(\bm{y}_{Q}^k)^T]^T=\bm{W}\bm{h}^k+\tilde{\bm{n}}^k\in \mathbb{C}^{QN_{RF}\times 1},
\end{equation}
where $\bm{W}=[\bm{W}_1^T,\cdots,\bm{W}_{Q}^T]^T\in \mathbb{C}^{QN_{RF}\times N}$, whose elements are randomly selected from the set $\frac{1}{\sqrt{N}}\{+1,-1\}$ with equal probability assuming that low-cost one-bit phase shifters are used\cite{beam_squint_LISTA}. The effective noise, $\tilde{\bm{n}}^k=[(\bm{W}_1\bm{n}_{1}^k)^T,\cdots,(\bm{W}_{Q}\bm{n}_{Q}^k)^T]^T$, is colored with the covariance matrix being $\bm{R}=\mathbb{E}\{\tilde{\bm{n}}^k(\tilde{\bm{n}}^k)^H\}=\mathrm{Blkdiag}(\sigma^2\bm{W}_1\bm{W}_1^H,\cdots,\sigma^2\bm{W}_{Q}\bm{W}_{Q}^H)$. To enhance the performance of CS algorithms, a pre-whitening procedure is adopted. Specifically, $\bm{R}$ is decomposed by Cholesky factorization as $\bm{R}=\sigma^2\bm{DD}^H$. Then, the pre-whitened received signal at the $k$-th subcarrier is
\begin{equation}
	\bm{\overline{y}}^k=\bm{D}^{-1}\bm{y}^k=\bm{\overline{W}}\bm{h}^k+\bm{\overline{n}}^k,
\end{equation}
where $\bm{\overline{W}}=\bm{D}^{-1}\bm{W}$, and $\bm{\overline{n}}^k=\bm{D}^{-1}\tilde{\bm{n}}^k$ is white. Eventually, concatenating the received signals, channels, and noises at all $K$ subcarriers column-wisely, we have
\begin{equation}
\bm{Y}=\bm{W}^H\bm{H}+\bm{N}.
\label{signal_model_matrix}
\end{equation}

\subsection{MmWave Wideband Channel Model}
In this paper, the clustered mmWave channel model\cite{sbl_unfolding_ce_frequency} is adopted. Considering half-wavelength antenna spacing, the response vector of an $N$-antenna ULA can be defined as 
\begin{equation}
	\bm{a}_N(\cdot)\triangleq[1,e^{-j\pi(\cdot)},\cdots,e^{-j\pi(N-1)(\cdot)}]^T/N,
\end{equation}
then the $k$-th uplink subchannel can be expressed as
\begin{equation}
\bm{h}^k=\sqrt{\frac{N}{N_cN_p}}\sum_{i=1}^{N_c}\sum_{j=1}^{N_p}\alpha_{i,j}e^{-j2\pi f_k \tau_{i,j}}\bm{a}_N(\psi_{i,j,k}),
\label{channel_model}
\end{equation}
where $N_c$ and $N_p$ denote the number of clusters and the number of subpaths in a cluster, respectively, and $f_k=f_c+(k-1-\frac{K-1}{2})\eta$ is the frequency of the $k$-th subcarrier with $f_c$ and $\eta=\frac{f_s}{K}$ denoting the central frequency and the subcarrier frequency spacing, respectively. Besides, $\alpha_{i,j}$, $\tau_{i,j}$, and $\psi_{i,j,k}$ denote the path gain, the delay, and the equivalent angle of arrival (AoA) at the BS of the $j$-th subpath in the $i$-th cluster, respectively. Due to the beam squint effect, $\psi_{i,j,k}=\frac{f_k}{f_c}\mathrm{sin}(\theta_{i,j})$ changes with $k$, where $\theta_{i,j}$ denotes the actual physical path angle. Detailed derivations about beam squint can be found in \cite{wideband_effect}. In the $i$-th cluster, we have $\theta_{i,j}=\bar{\theta}_i+\triangle\theta_{i,j}$ and $\tau_{i,j}=\bar{\tau}_i+\triangle\tau_{i,j},\forall j$, where $\bar{\theta}_i$ and $\bar{\tau}_i$ denote the mean AoA and mean delay, respectively, while $\triangle\theta_{i,j}$ and $\triangle\tau_{i,j}$ follow zero-mean Laplacian distributions\cite{sbl_unfolding_ce_frequency} with standard deviations $\sigma_{\theta_i}$ and $\sigma_{\tau_i}$, respectively. To express the wideband channel matrix neatly, we define the equivalent path gain as $\overline{\alpha}_{i,j}\triangleq \alpha_{i,j}e^{-j2\pi f_1\tau_{i,j}},\forall i,j$. Then, we have
\begin{equation}
\bm{H}\!=\!\sqrt{\frac{N}{N_cN_p}}\!\sum_{i=1}^{N_c}\!\sum_{j=1}^{N_p}\!\overline{\alpha}_{i,j}\!\left(\bm{a}_N\left(\mathrm{sin}(\theta_{i,j})\right)\!\bm{a}_K(2\eta\tau_{i,j})^T\right)\!\odot \bm{\Theta}(\theta_{i,j})
\end{equation}
where $\bm{a}_K(2\eta\tau_{i,j})$ can be viewed as the frequency-domain response vector and $\bm{\Theta}(\theta_{i,j})_{n,k}=e^{-j\pi n\mathrm{sin}(\theta_{i,j})(k-1-\frac{K-1}{2})\eta/f_c}$\cite{wideband_effect}. Notice that, when $N$ or $f_s$ is small, the beam squint effect can be ignored and $\bm{\Theta}$ is close to an all-one matrix.

\subsection{Problem Formulation}
Since the received signal is in the antenna-frequency domain while the channel is sparse in the AF and AD domains, proper dictionaries are required to realize domain transformations. For the delay dictionary, the commonly used oversampled DFT matrix defined as $\bm{A}_D\triangleq[\bm{a}_K(\varphi_1),\cdots,\bm{a}_K(\varphi_{G_D})]\in \mathbb{C}^{K\times G_D}$ is adopted, where the $G_D$ delay grids are 
\begin{equation}
\varphi_i=-1+(2i-1)/G_D,i=1,\cdots,G_D.
\end{equation}

Such a design, however, is not suitable for the angular dictionary with beam squint. Otherwise, the common angular sparsity structure in the AF domain will be destroyed and the channel sparsity in the AD domain will be reduced by energy diffusion\cite{beam_squint_ce}, leading to poor estimation performance. To compensate for beam squint, we adopt frequency-dependent angular dictionaries, where the angular dictionary for the $k$-th subchannel is defined as $\bm{A}^k_A\triangleq[\bm{a}_N(\phi^k_1),\cdots,\bm{a}_N(\phi^k_{G_A})]\in \mathbb{C}^{N\times G_A}$, and the $G_A$ angular grids are
\begin{equation}
	\phi^k_i=\frac{f_k}{f_c}(-1+(2i-1)/G_A),i=1,\cdots,G_A.
\end{equation}

With the pre-distortion factor $\frac{f_k}{f_c}$, the common angular sparsity structure in the AF domain still holds after experiencing the distortion of beam squint, and the energy diffusion in the AD domain is avoided. Denote the AF channel matrix and the AD channel matrix as $\bm{Q}\in \mathbb{C}^{G_A\times K}$ and $\bm{X}\in \mathbb{C}^{G_A\times G_D}$, respectively, we have $\bm{H}_{.k}=\bm{A}^k_A\bm{Q}_{.k},\forall k$ and $\bm{Q}=\bm{XA}_D^T$ if neglecting the quantization error, where $\bm{H}_{.k}$ denotes the $k$-th column vector of matrix $\bm{H}$. Vectorize all matrices in (\ref{signal_model_matrix}), we get the following transmission model in the standard CS form 
\begin{equation}
	\bm{y}=\bm{\Phi}\bm{x}+\bm{n},
	\label{signal_model_vector}
\end{equation}
\begin{equation}
	\bm{\Phi}\triangleq (\bm{I}_K\otimes \bm{W}^H)\bm{A}_A(\bm{A}_D\otimes \bm{I}_{G_A})\in \mathbb{C}^{KQN_{RF}\times G_DG_A},
	\label{Phi}
\end{equation}
where $\bm{\Phi}$ is called the measurement matrix, $\bm{A}_A\triangleq \mathrm{Blkdiag}(\bm{A}^1_A,\cdots,\bm{A}^K_A)$, and the covariance matrix of $\bm{n}$ is $\sigma^2\bm{I}_{KQN_{RF}}$. In a word, the goal of channel estimation is to accurately recover $\bm{x}$ based on $\bm{y}$, $\bm{\Phi}$, and $\sigma^2$. After the sparse AD channel is estimated, the original antenna-frequency channel can be readily reconstructed with the above dictionaries.

\begin{figure}[htb!] 
	\centering 
	\vspace{-0.35cm} 
	\subfigtopskip=2pt 
	\subfigbottomskip=1pt 
	\subfigcapskip=-3pt 
	\subfigure[With the common angular dictionary.]{
		\label{channel1}
		\includegraphics[width=0.45\textwidth]{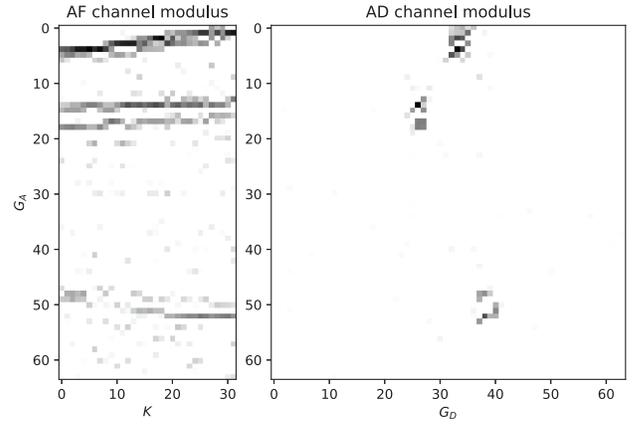}} 
	\subfigure[With frequency-dependent angular dictionaries.]{
		\label{channel2}
		\includegraphics[width=0.45\textwidth]{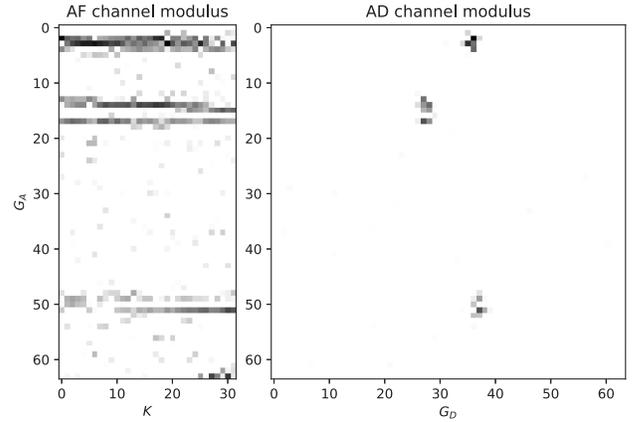}}
	\caption{An example of the estimated AF and AD channels' modulus with different angular dictionaries. In these images, the deeper the color, the larger the modulus.}
	\label{channel}
\end{figure}

To promote intuitive understanding of the beam squint effect and the impact of angular dictionaries, an example of modulus of the AF channel and the AD channel estimated by SBL is given in Fig. \ref{channel}, where the system and channel parameters are the same as the default setting in simulation. We can see that compared to using the oversampled DFT matrix as the common angular dictionary, using the adopted frequency-dependent angular dictionaries leading to aligned angular sparse supports of subchannels and sparser AD channel.

\section{AMP-SBL Unfolding-Based Channel Estimator}
In this section, the principles of the SBL algorithm are first introduced briefly. Then, two modifications made to lower its complexity and enhance its capability are elaborated in detail, leading to the AMP-SBL unfolding-based channel estimator.
\subsection{Principles of SBL}
\begin{figure}[htb!]
	\centering
	\includegraphics[width=0.5\textwidth]{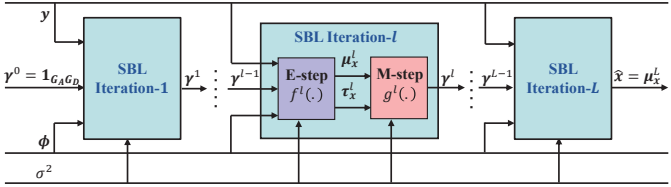}
	\caption{The SBL unfolding framework.}
	\label{SBL}
\end{figure}

As one of the most powerful CS algorithms, SBL has good sparse recovery performance and sufficient flexibility to exploit various sparsity structures\cite{sbl_ce}. To recover the sparse AD channel $\bm{x}$ in (\ref{signal_model_vector}), SBL assumes $\bm{x}$ follows the complex Gaussian distribution with a diagonal covariance matrix:
\begin{equation}
\bm{x}\sim \mathcal{CN}(\bm{0},\mathrm{diag}(\bm{\gamma})),
\end{equation}
where $\bm{\gamma}=[\gamma_1,\cdots,\gamma_{G_AG_D}]^T$ denote the variance parameters of the elements of $\bm{x}$. As illustrated in Fig. \ref{SBL}, after initializing $\bm{\gamma}^0$ as $\bm{1}_{G_AG_D}$, $L$ SBL iterations are executed where each iteration includes the expectation step (E-step) and the maximization step (M-step). Specifically, in the $l$-th iteration, the posterior mean and covariance are updated through the E-step, while the variance parameters, $\bm{\gamma}^l$, are updated through the M-step. Thanks to the sparse-promoting feature of SBL\cite{sbl}, the converged posterior mean at the last iteration will be a sparse vector, which is regarded as the channel estimation, i.e., $\hat{\bm{x}}=\bm{\mu}_{\bm{x}}^L$. Denote the functions executed in the E-step and the M-step in the $l$-th iterations as $f^l(\cdot)$ and $g^l(\cdot)$, respectively. In the original SBL algorithm\cite{sbl}, $f^l(\cdot)$ includes operations in (\ref{a_posterior_mean}) and (\ref{a_posterior_var}) while $g^l(\cdot)$ executes the operation in (\ref{update_gamma}): 
\begin{equation}
	\bm{\mu}_{\bm{x}}^l=\bm{R}_{\bm{x}}^{l-1}\bm{\Phi}^H(\bm{\Phi}\bm{R}_{\bm{x}}^{l-1}\bm{\Phi}^H+\sigma^2\bm{I})^{-1}\bm{y},
	\label{a_posterior_mean}
\end{equation}
\begin{equation}
	\bm{\tau}_{\bm{x}}^l\!=\!\mathrm{diag}\!\left(\bm{R}_{\bm{x}}^{l-1}\!-\!\bm{R}_{\bm{x}}^{l-1}\bm{\Phi}^H\!(\bm{\Phi}\bm{R}_{\bm{x}}^{l-1}\bm{\Phi}^H+\sigma^2\bm{I})^{-1}\bm{\Phi}\bm{R}_{\bm{x}}^{l-1}\right),
	\label{a_posterior_var}
\end{equation}
\begin{equation}
	\bm{\gamma}^l=|\bm{\mu}_{\bm{x}}^l|^2+\bm{\tau}_{\bm{x}}^l.
	\label{update_gamma}
\end{equation}

Detailed derivations of (13-15) can be found in \cite{sbl}. Next, modifications made to $f^l(\cdot)$ and $g^l(\cdot)$ will be introduced.

\subsection{Low-Complexity AMP-Based E-Step}
In spite of the superior performance, the original E-step of SBL has high complexity due to matrix inversions and multiplications, especially in AD estimators where the sparse vector and measurement matrix have large dimensions. In \cite{UAMP_SBL}, a low-complexity alternative implementation of the E-step is proposed, which is based on the AMP algorithm. With proper approximations, the most computational intensive part in the E-step can be obtained through several cheap matrix-vector multiplications. However, as the price of lower complexity, AMP-SBL is less robust to difficult measurement matrices that are not i.i.d Gaussian. Unitary preprocessing can help to alleviate this issue\cite{UAMP_SBL}. Performing singular value decomposition on $\bm{\Phi}$, we have $\bm{\Phi}=\bm{U\Sigma V}$, where $\bm{U}$ is a unitary matrix and does not change the noise covariance. Then, (\ref{signal_model_vector}) is transformed to 
\begin{equation}
\bm{r}=\bm{A}\bm{x}+\bm{w},
\end{equation}
where $\bm{r}=\bm{U}^H\bm{y}$, $\bm{A}=\bm{U}^H\bm{\Phi}$, and $\bm{w}=\bm{U}^H\bm{n}$. Applying the AMP-based E-step, the operations executed by $f^l(\cdot)$ are given by {\bfseries Algorithm 1}. Still, $\bm{\gamma}^0$ is initialized as $\bm{1}_{G_AG_D}$, while $\bm{\mu}_{\bm{x}}^0$ and $\bm{s}^{0}$ are initialized as $\bm{0}_{G_AG_D}$ and $\bm{0}_{KQN_{RF}}$, respectively. Notations $(\cdot)^{.2}$ and $./$ denote element-wise squaring and division, respectively.

\begin{algorithm}
	\caption{The E-step in the $l$-th iteration of AMP-SBL}
	\begin{algorithmic}[1]
		\State $\bm{\tau}_p = |\bm{A}|^{.2}\bm{\tau}_{\bm{x}}^{l-1}$;
		\State $\bm{p}=\bm{A}\bm{\mu}_{\bm{x}}^{l-1}-\bm{\tau}_p\cdot \bm{s}^{l-1},\ \bm{\tau}_s=\bm{1}./(\bm{\tau}_p+\sigma^2\bm{1})$;
		\State $\bm{s}^l=\bm{\tau}_s\cdot(\bm{r}-\bm{p}),\ \bm{\tau}_q = \bm{1}./(|\bm{A}^H|^{.2} \bm{\tau}_s$);
		\State $\bm{q}=\bm{\mu}_{\bm{x}}^{l-1}+\bm{\tau}_q\cdot(\bm{A}^H\bm{s}^l)$;
		\State $\bm{\mu}_{\bm{x}}^l=\bm{q}./(\bm{1}+\bm{\tau}_q\cdot \bm{\gamma}^{l-1}),\ \bm{\tau}_{\bm{x}}^{l}=\bm{\tau}_q./(\bm{1}+\bm{\tau}_q\cdot \bm{\gamma}^{l-1})$;
	\end{algorithmic}
\end{algorithm}

\subsection{DNN-Based M-Step}
The M-step in (\ref{update_gamma}) is optimal only if the elements of $\bm{x}$ are independent with each other, which is not true in practical sparsified clustered mmWave channel. Besides, SBL with this kind of M-step usually requires dozens of iterations to converge, thus resulting in high complexity. Last but not least, although the unitary preprocessing helps improve the robustness of AMP-SBL to some extent, it still cannot cope with the highly structured measurement matrix in (\ref{Phi}). In simulation, we find that AMP-SBL diverges as the iteration progresses in most cases. Motivated by the above limitations of the original M-step, we propose to use a DNN to learn the optimal M-step, whose variance parameter update rule enjoys both effectiveness and efficiency in real channel data\cite{sbl_unfolding_ce_frequency}. 

\begin{figure}[htb!]
	\centering
	\includegraphics[width=0.5\textwidth]{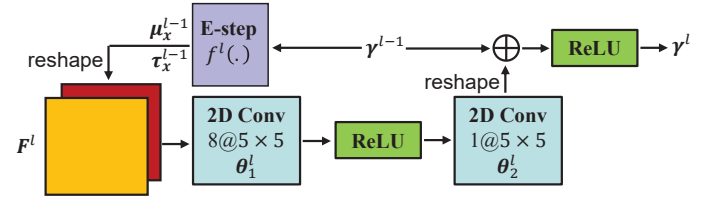}
	\caption{The architecture of the DNN in the $l$-th iteration.}
	\label{network}
\end{figure}

Since the channel sparsity and sparsity structures are reflected in the AD domain, we first reshape the features in the $l$-th iteration, $\bm{\mu}_{\bm{x}}^{l}$ and $\bm{\tau}_{\bm{x}}^{l}$, into $(G_A,G_D)$-dimensional matrices, and stack them along the expanded dimension to obtain the $(G_A,G_D,2)$-dimensional image-like input feature tensor, which is denoted by $\bm{F}^l$. Then, 2D convolutional (Conv) layers are naturally used to exploit the local correlation patterns in $\bm{F}^l$, as shown in Fig. \ref{channel}, and predict the variance parameter matrix, which is then reshaped back to the vector form for the computation of the next iteration's features. To improve training efficiency and stability, the residual connection is applied so that the DNN in each iteration learns the correction to the output variance parameters of the previous iteration. Through cross validation, we determine to use two Conv layers with $8$ and $1$ filters, respectively. The size of all filters is $3$. The DNN architecture in the $l$-th iteration is illustrated in Fig. \ref{network}, whose mapping function can be expressed as
\begin{equation}
	\bm{\gamma}^l=g_R(\bm{\gamma}^{l-1}+f^l_{C_2}(g_R(f^l_{C_1}(\bm{F}^l;\bm{\theta}^l_1));\bm{\theta}^l_2)),
	\label{update_gamma_uamp_residual}
\end{equation}
where $f^l_{C_1}(\cdot;\bm{\theta}^l_1)$ and $f^l_{C_2}(\cdot;\bm{\theta}^l_2)$ denote the convolution operations of the first and the second Conv layers with weights $\bm{\theta}^l_1$ and $\bm{\theta}^l_2$, respectively, while $g_R(x)=\mathrm{max}(0,x)$ denotes the ReLU activation function to introduce non-linearity and guarantee non-negativity.

We generate $8,000$, $1,000$, $1,000$ channel samples for network training, validation, and testing, respectively. Batch size is set to $128$, and random noises are re-generated for each batch to increase dataset diversity. The optimizer is Adam and the loss function is mean-squared error (MSE). Layer-wise training is adopted, which is critical for the convergence of the AMP-SBL unfolding algorithm. Firstly, a two-iteration residual network is constructed, where the weights of convolution filters are randomly initialized and trained to convergence. After that, new iterations are added one at a time, where the weights of the DNN in the newly added iteration are copied from the previous iteration. The entire network is then trained to convergence. In this way, totally $L-1$ times of training are required. Same strategies are used in each training, where the initial learning rate is set to $10^{-3}$ and decays with a factor of 10 and a patience of 4 epochs to accelerate training, and early stopping with a patience of 10 is used to avoid overfitting. 

\subsection{Baseline Algorithms and Complexity Analysis}
The numbers of real floating operations (FLOPs) of different algorithms are listed in Table I. For notation simplicity, we define $G\triangleq G_AG_D$ and $M\triangleq QN{_RF}$. Compare the first three algorithms in the Table, we can see that the extra FLOPs of the DNN-based M-step is marginal, while the AMP-based E-step has significantly lower complexity than the original implementation. Thanks to smaller dimensions of measurement matrices and sparse vectors, AF estimators have lower complexity than AD estimators. However, their performance is not satisfactory, as will be shown in simulation. Apart from the original SBL, heuristic variants including PC-SBL\cite{pc_sbl} and M-SBL\cite{m_sbl} are also compared, which can exploit the AD block sparsity structure and the AF common sparsity structure to improve performance without complexity overhead. We also reproduce the LISTA\cite{beam_squint_LISTA} algorithm, which uses deep learning and considers beam squint. Furthermore, the FLOPs of reconstructing the original channel from the estimated sparse channel should be included, whose expressions are $8KG_AN$ and $8KG_AN+8KG$ in AF and AD estimators, respectively. To show the impact of dictionary, results with the common frequency-independent angular dictionary (FID) will also be given in simulation, in which case only M-SBL AF's per-iteration FLOPs changes to $16M^2G_A+8KMG_A$.
\begin{table}[!htb]
	\begin{tabular}{|c|c|}
		\hline
		Algorithm & FLOPs per Iteration \\ \hline
		(PC-)SBL\cite{pc_sbl} & $16(KM)^2G$\\ \hline
		SBL unfolding\cite{sbl_unfolding_ce_frequency} & $(16(KM)^2+432)G$\\ \hline
		AMP-SBL unfolding & $(20KM+432)G$\\ \hline
		(M-)SBL AF\cite{m_sbl} & $16KM^2G_A$ \\ \hline
		LISTA\cite{beam_squint_LISTA} & $4K\left(\left(4M+256\right)N+32768\right)$ \\ \hline
	\end{tabular}
	\centering
	\caption{FLOPs per iteration of different algorithms.}
	\label{complexity}
\end{table}

\section{Simulation Results}
\label{simulation}
In this section, simulation results are provided to validate the superiority of the proposed approach. The default system and channel parameters are as follows: $N=32$, $N_{RF}=4$, $Q=4$, $N_c=3$, $N_p=10$, $\alpha_{i,j}\sim \mathcal{CN}(0,1),\bar{\theta}_i\sim \mathcal{U}[0,2\pi],\bar{\tau}_i\sim \mathcal{U}[0,25\,\text{ns}],\sigma_{\theta_i}=4^\circ,\sigma_{\tau_i}=0.06\,\text{ns}$, $f_c=28$ GHz, $f_s=4$ GHz, $K=32$, $G_A=G_D=64$, $\text{SNR}=10$ dB. NMSE defined as $\mathbb{E}\{||\bm{H}-\hat{\bm{H}}||_F^2/||\bm{H}||_F^2\}$ is utilized to measure the channel estimation performance, which is obtained by averaging 200 random data samples. For reproduction, the source code is available at \href{https://github.com/EricGJB/Angular_Delay_SBL_Unfolding}{$\mathrm{https://github.com/EricGJB/AD\_SBL\_Unfold}$}.

\begin{figure}[htb!]
	\centering
	\includegraphics[width=0.45\textwidth]{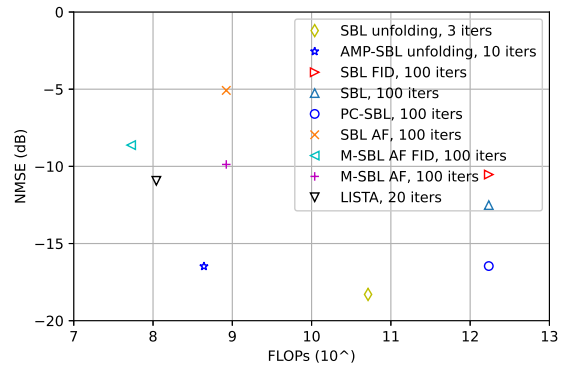}
	\caption{Performance versus complexity.}
	\label{tradeoff}
\end{figure}

Fig. \ref{tradeoff} illustrates the performance-complexity tradeoffs of different algorithms, which are vividly reflected by their positions on the FLOPs-NMSE plane. All algorithms are executed iteratively until the performance converges, and the numbers of iterations are shown in the legend. We can see that the performance of AF estimators is poor, which validates the motivation of using AD estimators. Although PC-SBL and SBL unfolding have similar or even better performance than AMP-SBL unfolding, their complexity is much higher due to the more complex E-step and slower convergence, making them more like performance limits than practical choices. LISTA has very low complexity, but its performance is not satisfactory. Therefore, to some extent, we can claim that the proposed approach located in the lower left corner of the plane achieves the best performance-complexity tradeoff. Besides, the advantage of frequency-dependent angular dictionaries is demonstrated as well, which comes from the improved AF channel sparsity structure and AD channel sparsity illustrated in Fig. \ref{channel}. Notice that, AMP-SBL is not in the figure since it fails with very large NMSE, which means the embedding of DNNs is more than beneficial, but necessary to enable the low-complexity AD estimator, especially with difficult measurement matrices. 

\begin{figure}[htb!]
	\centering
	\includegraphics[width=0.45\textwidth]{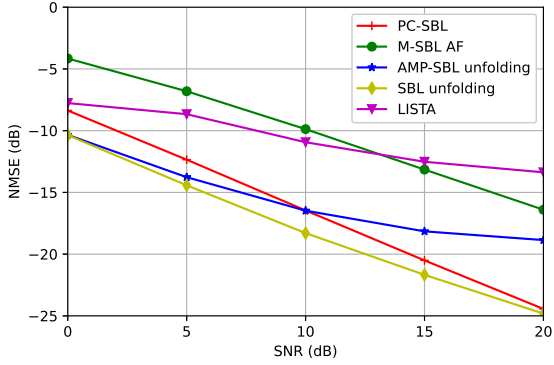}
	\caption{Performance versus SNR.}
	\label{impact_of_snr}
\end{figure}

To verify the generality of the proposed approach, Fig. \ref{impact_of_snr} illustrates the impact of SNR, where $Q=4$. As we can see, AMP-SBL unfolding consistently outperforms M-SBL AF and LISTA, and achieves almost the same performance as SBL unfolding in low SNR regimes, demonstrating its superiority. In high SNR regimes, the performance of AMP-SBL unfolding meets plateau and becomes inferior to PC-SBL and SBL unfolding due to the approximations made in the E-step. Nevertheless, the extremely low complexity makes AMP-SBL unfolding very appealing for practical applications even with only moderate performance. 

\begin{figure}[htb!]
	\centering
	\includegraphics[width=0.45\textwidth]{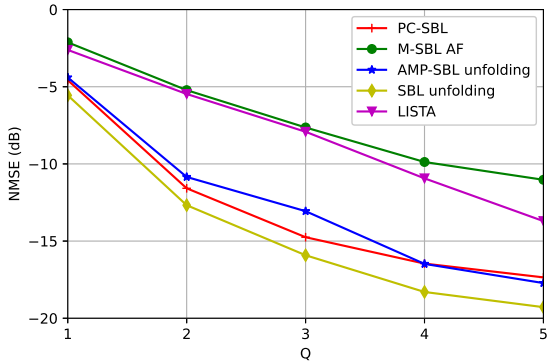}
	\caption{Performance versus $Q$.}
	\label{impact_of_Q}
\end{figure}

The impact of $Q$ is also illustrated in Fig. \ref{impact_of_Q}, where $\text{SNR}=10$ dB. As we can see, AMP-SBL unfolding achieves similar performance to PC-SBL and much better performance than M-SBL AF and LISTA with various $Q$, which also results in its superiority in terms of pilot overhead saving. For instance, the NMSE of AMP-SBL unfolding with $Q=2$ is comparable to the NMSE of LISTA with $Q=4$ and the NMSE of M-SBL AF with $Q=5$, therefore saving twice and three times the pilot overhead, respectively.

\section{Conclusion}
In this paper, we first compensate for beam squint by using frequency-dependent angular dictionaries. Then, we have proposed an AMP-SBL unfolding-based approach for angular-delay domain wideband mmWave massive MIMO channel estimation. Compared to the original SBL algorithm, the AMP-based E-step reduces complexity significantly, while the DNN-based M-step enables accurate estimation with the highly structured measurement matrix. Simulation results demonstrate the good performance-complexity trade-off of the proposed approach with various system configurations. We will also consider to extend to the more challenging near-field terahertz scenario in the future.

\end{document}